\documentclass[twocolumn,showpacs,preprintnumbers,amsmath,amssymb]{revtex4}
\usepackage{graphicx}
\usepackage{dcolumn}
\usepackage{bm}

\def\bd{\begin{document}} \def\ed{\end{document}}
\def\bmp{\begin{minipage}} \def\emp{\end{minipage}}
\def\bcc{\begin{center}} \def\ecc{\end{center}}     \def\npg{\newpage}
\def\beq{\begin{equation}} \def\eeq{\end{equation}} \def\hph{\hphantom}
\def\be{\begin{equation}} \def\ee{\end{equation}} \def\r#1{$^{[#1]}$}
\def\n{\noindent} \def\ni{\noindent} \def\pa{\parindent}
\def\hs{\hskip} \def\vs{\vskip} \def\hf{\hfill} \def\ej{\vfill\eject}
\def\cl{\centerline} \def\ob{\obeylines}  \def\ls{\leftskip}
\def\underbar#1{$\setbox0=\hbox{#1} \dp0=1.5pt \mathsurround=0pt
   \underline{\box0}$}   \def\ub{\underbar}    \def\ul{\underline}
\def\f{\left} \def\g{\right} \def\e{{\rm e}} \def\o{\over} \def\d{{\rm d}}
\def\vf{\varphi} \def\pl{\partial} \def\cov{{\rm cov}} \def\ch{{\rm ch}}
\def\la{\langle} \def\ra{\rangle} \def\EE{e$^+$e$^-$} \def\pt{p_{\rm t}}
\def\bitz{\begin{itemize}} \def\eitz{\end{itemize}}
\def\btbl{\begin{tabular}} \def\etbl{\end{tabular}}
\def\btbb{\begin{tabbing}} \def\etbb{\end{tabbing}}
\def\beqar{\begin{eqnarray}} \def\eeqar{\end{eqnarray}}
\def\\{\hfill\break} \def\dit{\item{-}} \def\i{\item}
\def\bbb{\begin{thebibliography}{9} \itemsep=-1mm}
\def\ebb{\end{thebibliography}} \def\bb{\bibitem}
\def\bpic{\begin{picture}(260,240)} \def\epic{\end{picture}}
\def\akgt{\noindent{Acknowledgements}}
\def\fgn{\noindent{\bf\large\bf Figure captions}}
\bd

\title{A comparison of different methods\\
in the study of dynamical fluctuations\\
in high energy \EE collisions}

\thanks{This work is supported in part by NSFC under project
19975021.}

\author{Chen Gang$^{1, 2}$  and Liu Lianshou$^1$}

\affiliation{1 \ Institute of Particle Physics, Huazhong 
Normal University, Wuhan 430079 China\\
2 \ Department of Physics, Jingzhou Teacher's College,
 Hubei 434104 China}

\begin{abstract}
Different methods in the study of anomalous scaling of factorial
moments in high energy \EE collisions are examined in some detail. 
It is shown that the horizontal and vertical factorial moments are
equivalent only when they are used in combination with the cumulant 
variables. The influence of different reference frames and that of
phase space restrictions is also discussed.
\end{abstract}


\maketitle
\section{Introduction}

Since the observation of exotic multiparticle events in Cosmic Ray
experiments~\cite{CR} and especially the discovery of unexpectedly 
large local fluctuations recorded by the JACEE collaboration~\cite{JACEE}, 
the investigation of nonlinear phenomena (NLP)
in high energy collisions has attracted much attention~\cite{Kittel}.
One of the signals of these NLP is the fractal 
property~\cite{Mandelbrote} of
the multiparticle final states in high energy collisions.
This property can be characterized by the anomalous scaling of the 
probability moments defined as
\beqar 
 C_q(M)=\frac{1}{M}\sum_{m=1}^M \frac{\la p_m^q \ra }{\la p_m\ra^q},
\eeqar
where a certain phase space region $\Delta$ is divided in a proper way 
(isotropically for a self-similar fractal while anisotropically for
a self-affine fractal~\cite{FLPRL}) into $M$ sub-cells,
$p_m$ is the probability for a particle to fall into the $m$th sub-cell,
$\la \cdots \ra$ denotes the average over an event sample.
If the $C_q(M)$'s have a power law behaviour with 
the diminishing of phase space scale:
\beqar 
 C_q(M) \sim M^{\phi_q} \qquad (M\to \infty), 
\eeqar
then it is said to be anomalous scaling and the system is a fractal.

In real experiments the probability $p_m$ is unobservable and the 
corresponding moments $C_q$ is unaccesible. This problem has been solved
by Bia\l as and Peschanski~\cite{BP}, who are able to show that
the factorial moments 
\beqar 
 F_q(M) =
    \frac{1}{M}\sum\limits_{m=1}^{M}
    \frac{\la n_m(n_m-1) \cdots (n_m-q+1)\ra}
    {\la n_m \ra ^q  } 
\eeqar
are equal to the probability moments $C_q$ provided the statistical 
fluctuations are Poissonian. Thus the scaling property of factorial
moments, sometimes called intermittency, 
becomes a central problem in the study of nonlinear
phenomena in high energy collisions. 

Various methods have been developed in this study.

Firstly, people noticed that in the definition Eq.(3) of factorial moments
the average over event sample (vertical average) is carried out first and
then is the average over the $M$ sub-cells (horizontal average). It was proposed
to exchange the order of these two averages and define the horizontal
factorial moments (HFM)~\cite{HFM} as
\beqar 
 F_q^{(\rm H)}(M) = 
    \frac{\f\la {M^{-1}}\sum\limits_{m=1}^{M} n_m(n_m-1) \cdots (n_m-q+1)\g\ra}
    {\f\la {M^{-1}}\sum\limits_{m=1}^{M} n_m \g\ra ^q  } .
\eeqar
Accordingly, the $F_q$ defined in Eq.(3) is called vertical factorial moments 
(VFM).

Note that the equality of factorial moments $F_q$ and probability moments
$C_q$ has been proved only for the VFM. Therefore, in the study of the 
nonlinear phenomena --- fractal property of multiparticle system, the
HFM is appropriate only when it is equal to VFM. We will see in the following 
that this equality holds in some cases but does not hold in some other cases.

Secondly, various methods have been proposed to correct for the unflatness of
the phase-space variable distributions. One is to divide the factorial 
moments by a factor $R_q$~\cite{Fialkowski}
\beqar 
F_q^{\rm C} = \frac{F_q}{R_q}, 
  \qquad R_q = \frac{{M^{-1}} \sum\limits_{m=1}^{M} \la n_m\ra ^q} 
  {\f\la {M^{-1}}\sum\limits_{m=1}^{M} n_m \g\ra ^q  } .
\eeqar 
Another one is to change the phase space variable $x$ 
into the corresponding
cumulant variable $x_{\rm c}$ before calculating the factorial moments. The
cumulant variable is defined as~\cite{Cummulant}
\beqar 
 x_{\rm c}=\frac {\int_{x_{\rm min}}^{x} \rho (x) \d x}
{\int_{x_{\rm min}}^{x_{\rm max}} \rho (x) \d x},
\eeqar
where $x_{\rm min}$ and $x_{\rm max}$ are the two boundaries of the $x$ region,
respectively.

\begin{figure}
\includegraphics[width=4cm]{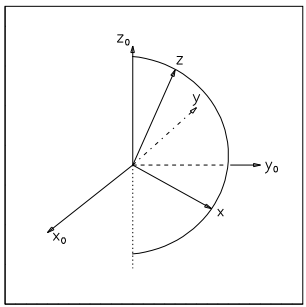}
\caption{\label{fig:rotate} The rotated coordinate system}
\end{figure}

\begin{figure}
\includegraphics[width=6.2cm]{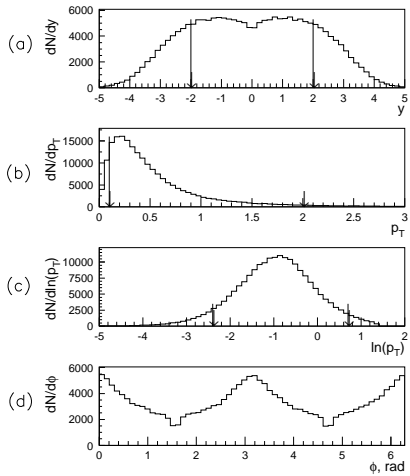}
\caption{\label{fig:distribution} Average distribution of phase space variables}
\end{figure}

Another problem arises while carrying on this kind of study in \EE collisions.
In these collisions the thrust (or sphericity) 
axis is chosen as the $z$ axis (longitudinal axis) to define the 
phase space variables: rapidity $y$, transverse momentum 
$\pt$ and azimuthal angle $\vf$. Different frames could be used to define 
the azimuthal angle $\vf$. The first one is to choose the minor axis of thrust 
(or sphericity) analysis as the $x$ axis, and use it as the starting point for 
counting the azimuthal angle $\vf$. 
The second one is to put the $z$ axis still on the major thrust axis,
but turn the coordinate system around it and let the new $x$ axis
lie on the $z_0$-$z$ plan\cite{LLL}, where $x_0,y_0,z_0$ denote the axes of
the lab system and $x,y,z$ those of the turned system, as shown in Fig.1.
In the following this frame will be referred to as the ``rotated frame''.
The third method is to rotate the frame in each event
for a random angle around the $z$ axis~\cite{CGDatong}.
This is called the random frame.

All these methods have been used in the literature for studying the
anomalous scaling of factorial moments, making the results hard
to be compared. In the present paper we will examine these methods
in some detail and discuss their applicability in physical problems.
We will take \EE collisions at the Z$^0$ energy $\sqrt s=91.2$ GeV as example
and use JETSET7.4~\cite{Jetset} 
Monte Carlo code to generate 500 000 multihadron events as the event sample.
 
\section{Average distributions of phase space variables}

In Fig.2 ($a$), ($b$) and ($d$) are shown the average distributions
of $y$, $\pt$ and $\vf$, where the rapidity is defined as
$y=0.5 \ln [(E+p_z)/(E-p_z)]$ with $z$ along the thrust axis; the
azimuthal angle $\vf$ is defined in the plane perpendicular to
the thrust axis, calculated with respect to the minor axis. 

It can be seen from the figures that all the distributions are unflat.
Especially, the distribution of $\pt$ is exponential and is highly
concentrated in low $\pt$. A simple variable transformation
$\pt \to \ln \pt$~\cite{OPAL} 
can make it a litter flatter as shown in Fig.2 ($c$).

\begin{figure}
\includegraphics[width=6.2cm]{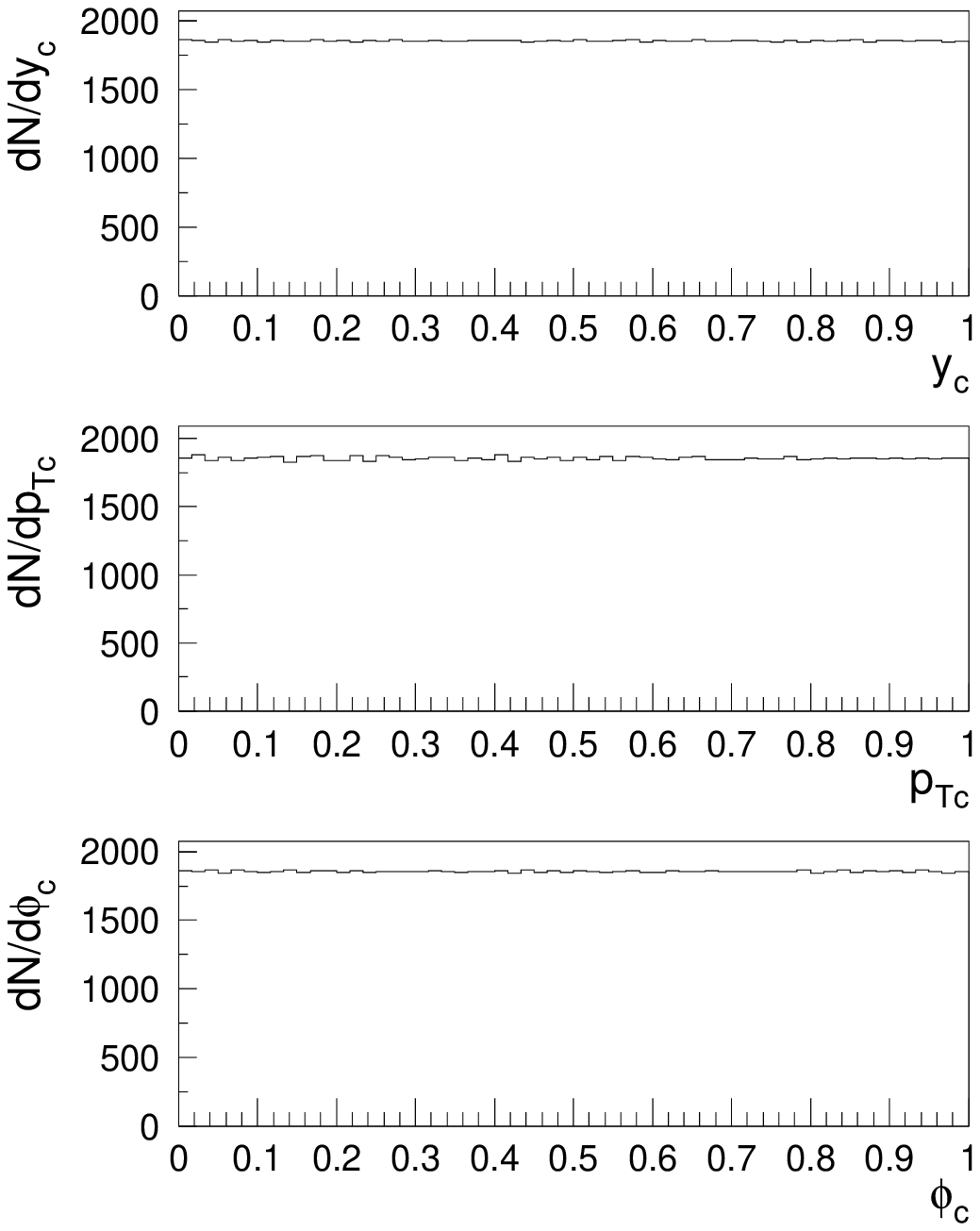}
\caption{\label{fig:cumulant} The distribution of cumulant variables}
\end{figure}

\begin{figure*}
\includegraphics[width=18cm]{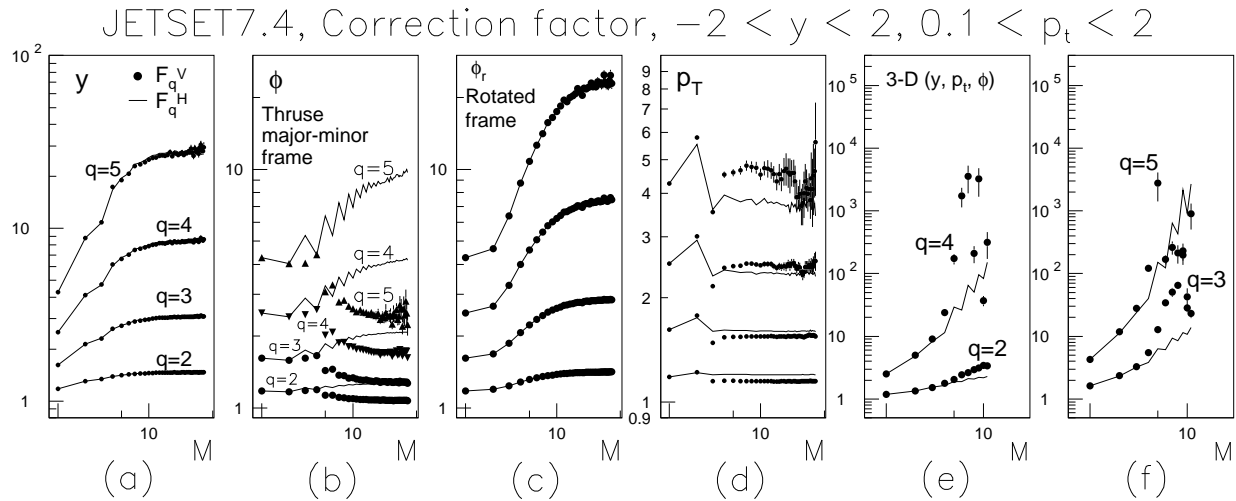}
\caption{\label{fig:correction}VFM (solid circles) and HFM (solid curves) using 
correction factor method\\ (restricted phase space).
In ($b$) the upward triangles are for $q=5$, downward ones for $q=4$.}
\end{figure*}
\begin{figure*}
\includegraphics[width=18cm]{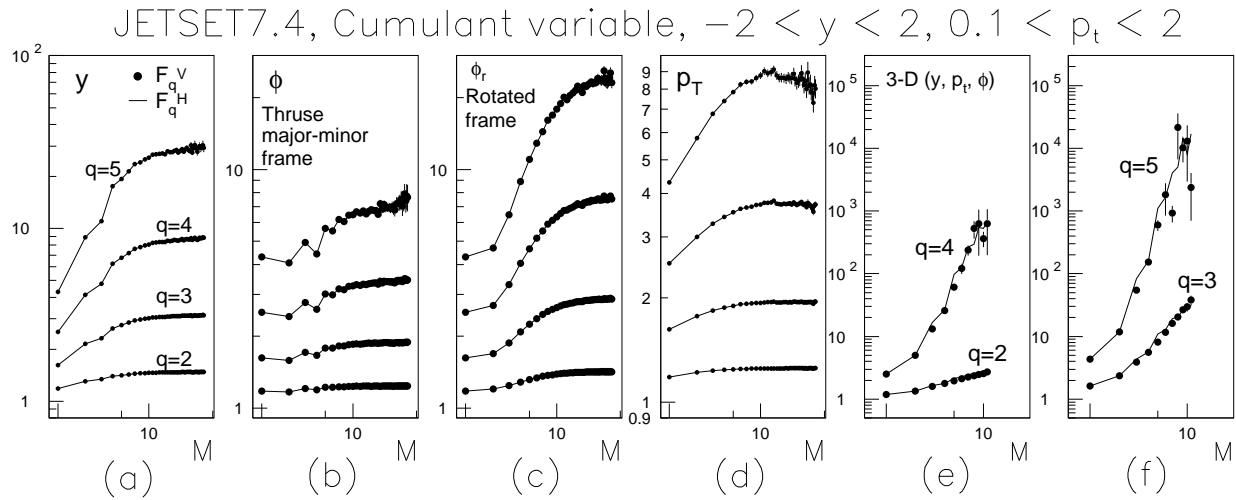}
\caption{\label{fig:restrict}VFM (solid circles) and HFM (solid curves) using 
cumulant variables\\ (restricted phase space)}
\end{figure*}

\section{The correction factor and cumulant variables}

The unflat average distribution will cause additional variation of factorial 
moments with the diminishing of phase space scale and make the scaling
property of factorial moments unequal to that of the probability moments
even for the VFM. This effect has to be corrected.

Fialkowski proposed a factor~\cite{Fialkowski}, 
cf. Eq.(5), to correct for this effect.  
This method works good when the distribution of the phase space variable
is not far from flat, e.g. the distribution of rapidity $y$ in a restricted 
central region $|y|<Y_{\rm c}$  
with $Y_{\rm c}=2$ as shown in Fig.2 ($a$), and is 
not good when the distribution is far from flat. This is especially
the case for the distribution of $\pt$, cf. Fig.2 ($b$). Therefore, people
sometimes transform $\pt\to\ln\pt$ first~\cite{OPAL} 
and calculate $F_q(\ln\pt)$ instead of $F_q(\pt)$, and then correct 
the result by the factor $R_q$ given in Eq.(5).
Note that the highly asymetric region $0.1 \leq \pt \leq 2$ is transformed
to the region $-2.3 \leq \ln\pt \leq 0.69$, which is distributed symmetrically
around the pick of distribution, cf. the two arrows in Fig's.2 ($b$) and ($c$).
 
Since a transformation to a flatter distribution is necessary before calculating
factorial moments, it is evident that the best way is to transform all the 
phase space variables to a flat distribution first. 
This could be established through the 
transformation to cumulant variables~\cite{Cummulant}, 
cf. Eq.(6). The corresponding distributions are shown in Fig.3.

\begin{figure*}
\includegraphics[width=18cm]{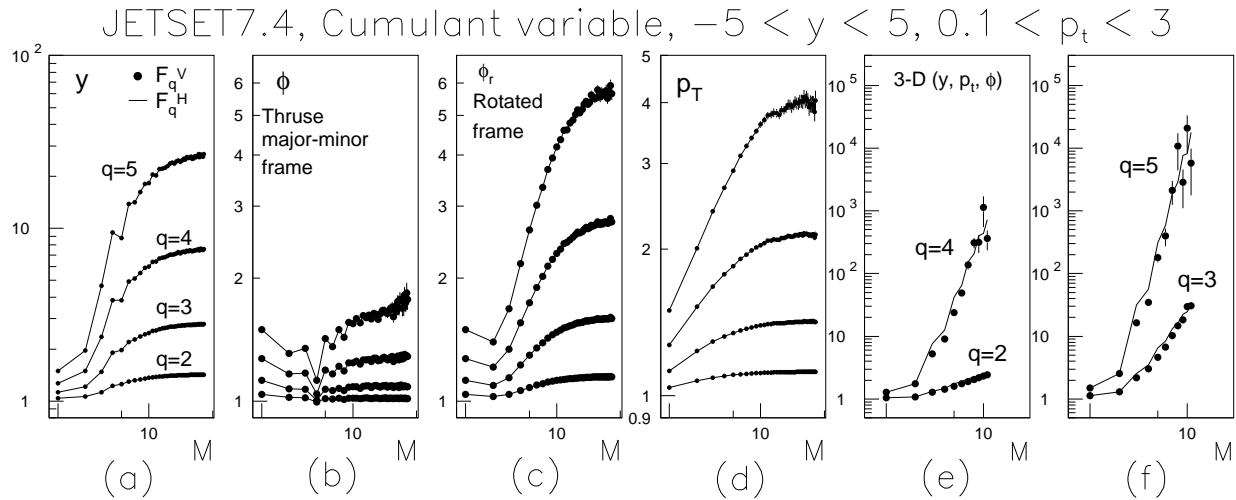}
\caption{\label{fig:fullspace}VFM (solid circles) and HFM (solid curves) using 
correction factor method\\ (nearly full phase space)}
\end{figure*}

\section{Vertical and horizontal factorial moments}

Now let us turn to the comparison of vertical and horizontal factorial moments
(VFM and HFM). 

As noticed in the Introduction, our aim is to explore the anomalous scaling
of probability moments as shown in Eq.(2) but the equivalence of factorial
and probability moments has been proved only for the VFM. So, the HFM is
appropriate only when it is equal to VFM. 

In Figures 4 and 5 $(a),(b),(d),(e),(f)$ 
are shown the 1-D and 3-D VFM (solid circles) and HFM 
(solid curves) for the moment orders $q=2,3,4,5$ calculated using the 
correction factor, Eq.(5), method and the cumulant variables Eq.(6), 
respectively.

It can be seen from the figures that the VFM and HFM are equal only when using
together with the cumulant variables and are unequal, especially for 1-D
$F_q(\vf)$ and $F_q(\pt)$, when using the correction factor method.

The results from VFM + Correction Factor method are about the same as that 
from VFM + Cummulant Variables for 1-D $F_q(y)$ and 3-D $F_q(y,\pt,\vf)$ but 
are not the case for 1-D $F_q(\vf)$ and $F_q(\pt)$, cf. Fig's. 4 and 5.

\section{The influence of phase space reduction}

In hadron-hadron and nucleus-nucleus collisions the central rapidity regions
with the rapidity $y$ restricted to $|y|\leq Y_{\rm c}$ for some value of
$Y_{\rm c}$ is commonly used. This is physically meaningful, because 
in these collisions the final state particles are mainly produced in the 
central region, while the particles in the regions $|y|>Y_{\rm c}$ are mainly 
come from the fragmentation of incident particles (leading particle effect).

On the contrary, in \EE collisions, the  multihadron
final state is produced from a
point source --- virtual photon or Z$^0$, and no leading particle effect
is present. The rapidity $y$ is usually defined with respect to the thrust
or sphericity axis. In this case, to carry on the study in a ``central
rapidity region'' $|y|\leq Y_{\rm c}$ is physically doubtful. This is especially
evident when the collision energy is so high that 3, 4 or even more jets
can be produced. In a 2-jet event the restriction $|y|\leq Y_{\rm c}$ will cut
out the most energetic particles from the two opposite jets symmetrically, but
in a 3-jet event the same cut will cut out the most energetic particles only 
from one jet while retain almost all the particles in the other two jets.
This asymmetric cut will results in unexpected phenomena, and the
physical meaning may be difficult to interpret. 
Therefore, the study of multiparticle dynamics in \EE collisions 
is better to be carried out in the full phase space.  
However, the central rapidity region is sometimes also used in the literature
for the study of \EE collisions~\cite{OPAL}. 
Therefore, to investigate the influence of rapidity cut is worthwhile.

In Fig.6 
are shown the results of VFM and HFM in nearly the full phase space
--- $-5 < y < 5, 0 < \vf < 2\pi, 0.1 < \pt < 3$ GeV, to be compared with
the results shown in Fig.5 for a restricted phase space ---
$-2 < y < 2, 0 < \vf < 2\pi, 0.1 < \pt < 2$ GeV. 
 
It can be seen from Fig.6 that the first point in 3-D $F_q(y,\pt,\vf)$ 
and 1-D $F_q(y)$
and the first 3 points in 1-D $F_q(\vf)$ do not lie on a scaling curve together
with the other points. This is due to the momentum conservation 
effect~\cite{MMCN}.  In the anomalous-scaling study, 
these points should be omitted.

The momentum conservation effect will also be reduced in a restricted phase
space region, which was first pointed out in Ref.\cite{MMCN} 
and has been proposed as a second
method for eliminating the influence of this effect. This explains
the reason why the first points in Fig's.5 lie on the scaling curves.

\section{The fluctuations in azimuthal angle}

The fluctuations in azimuthal angle are worthwhile special investigation. 
It is commonly expected that there should have 
cylindrical symmetry around the $z$ axis.
If that is the case, then the fluctuations should have no correlation with the
$x$ axis chosen for counting the azimuthal angle $\vf$. In Fig's. 4, 5, 6 ($b$)
are shown the $F_q(\vf)$ with $x$ axis along the minor of thrust analysis,
while in the corresponding figures ($c$) are shown the results after rotating
the $x$ axis to let it lie on the $z$-$z_0$ plane~\cite{LLL}, cf. Fig.1. 
It can be seen that in the rotated frame, Fig's. 4, 5, 6 ($c$),
the $F_q(\vf)$ increases much faster as the diminishing of 
phase space scale than that in the thrust-minor frame, Fig's. 4, 5, 6 ($b$).
As discussed in Ref.~\cite{LLL} this is because the thrust-minor axis is
basically determined by the first hard gluon emission and taking this
axis as $x$ axis to count the azimuthal angle $\vf$ will highly reduce the
fluctuation of the direction of first hard gluon emission. 
After rotation, the correlation between $x$ axis and 
the direction of first hard gluon emission is relaxed and the
full dynamical fluctuations are exhibited.

We could also rotate the $x$ axis around $z$ for a random angle in each
event~\cite{CGDatong}. 
The resulting $F_q(\vf)$ turn out to be the same as those in the 
rotated frame with $x$ on the $z$-$z_0$ plane shown in Fig's.4, 5, 6 ($c$). 
This confirms the cylindrical symmetry of the fluctuation in $\vf$ after the
correlation with the thrust-minor is relaxed.

\section{Conclusions}

The following conclusions could be drawn from the above investigation:

1) The horizontal factorial moments (HFM) are equivalent to the vertical ones
(VFM) only after the cumulant-variable transformation. Therefore, in the
study of non-linear phenomena (intermittency or fractal) in high energy 
collisions the HFM could be used only in combination with the cumulant 
variables. On the other hand, the HFM is in its own right
useful in single-event anaylsis.
It can be seen from Eq.(4) that HFM is the average of the so called
single-event factorial moments $f_q^{(\rm e)}$~\cite{HWA} 
\beqar  
 F_q^{(\rm H)}(M) &=& 
    {\la f_q^{(\rm e)}\ra } / {\la f_1^{(\rm e)}\ra^q}.\\
 f_q^{(\rm e)}(M) &=&
    \frac{1}{M}\sum\limits_{m=1}^{M} n_m^{(\rm e)}(n_m^{(\rm e)}-1) \cdots 
       (n_m^{(\rm e)}-q+1)
\eeqar
where $n_m^{(\rm e)}$ is the number of particles of a single event in the 
$m$th sub-cell. The fluctuation of the single-event factorial moments 
around its average (HFM) is a characteristic of single-event 
fluctuations~\cite{HWA}.

2) The scaling properties of factorial moments in transverse directions
($\vf, \pt$) are very sensitive to the correction method used. They are 
unstable when using the correction factor method, Eq.(5),
even after the transformation
$\pt\to\ln\pt$ has been made. Using this method, the VFM in $\vf$ and the HFM
in $\pt$ fall down instead of increase with the diminishing of phase space 
scale, while at the same time the HFM in $\vf$ and the VFM in $\pt$ do
increase with the diminishing of scale, cf. Fig's. 4 ($b$) and ($d$). 

3) In the full phase space, the first few points of factorial moments do 
not lie on the scaling curve with the other points, due to the 
momentum conservation effect. This effect can be eliminated either through 
neglecting these points or through a cut in phase space. 

4) The thrust (or sphericity) major-minor frame is inappropriate for the 
study of the scaling property of the azimuthal angle $\vf$, because this frame
is strongly correlated with the direction of first hard gluon emission.
Rotate the $x$ axis to let it 
lie on the $z$-$z_0$ plane or rotate it randomly 
for each event can relax this correlation and exhibit the full dynamical
fluctuations in $\vf$.

Therefore, the cumulant variables together with a frame rotated around
the thrust (or sphericity) 
axis is the best for the investigation of the nonlinear phenomena (anomalous 
scaling of probability moments) in \EE collisions. The VFM and HFM are
equivalent in this case.

\begin{acknowledgments}
The authors thank Wolfram Kittel and Hu Yuan for helpful discussions.
\end{acknowledgments}

\def\J#1#2#3#4{{\it #1} {\bf #2}, {#3} (#4).}
\def\PRL{Phys. Rev. Lett.} \def\PRep{Phys. Rep.}
\def\PRD{Phys. Rev. D} 
\def\NPB{Nucl. Phys. B}  \def\PLB{Phys. Lett. B}

\ed